


\magnification=1200
\newcount\sec
\sec=0
\newcount\dummy

\newcount\num
\def\eqnum{\global\advance\num by 1 \eqno(\the\sec.\the\num )}
\def\eqnlbl#1{\eqnum\xdef #1{(\the\sec.\the\num )}}
\def\eqnumapp#1{\global\advance\num by 1 \eqno({\rm #1}.\the\num )}
\def\eqnumapplbl #1#2 {\eqnumapp#1\xdef #2{({\rm #1}.\the\num )}}
\def\eqalnum{\global\advance\num by1 (\the\sec .\the\num)}
\def\eqalnumlbl#1{\eqalnum\xdef #1{(\the\sec.\the\num )}}

\newcount \refno
\refno=0
\xdef\gobble#1{}
\def\ifundefined #1
     {\expandafter\ifx\csname\expandafter\gobble\string#1\endcsname\relax}

\def\ref #1{$
   \ifundefined {#1}
   \global\advance\refno by 1
   \the\refno
   \xdef #1 {\the\refno}
   \else #1 \fi   $}


\def\writeref{
 \def\ref ##1{$
   \ifundefined {##1}
   \global\advance\refno by 1
   \the\refno
   \xdef ##1 {\the\refno\string ##1}
   \else ##1 \fi   $}     }

\font\Bigrm=cmr10 scaled\magstep2
\font\bigrm=cmr10 scaled\magstep1

\font\Bigit=cmti10 scaled\magstep2




\def\heading#1{\goodbreak\bigskip\line{\bf #1\hfil}
        \nobreak\smallskip\nobreak\noindent}

\def\leaderfill{\leaders\hbox to 1em{\hss.\hss}\hfill}

\def\half{{\scriptstyle{1\over 2}}}
\def\third{{\scriptstyle{1\over 3}}}
\def\quarter{{\scriptstyle{1\over 4}}}



\def\Z{\hbox{Z$\!\!$Z}}

\null
\line{\hfil SHEP-91/92-24 \quad}
\vfill
\line{\hfil
{\Bigrm Discrete Symmetries from Broken {\Bigit SU(N)} }
\hfil}
\line{\hfil
{\Bigrm and the MSSM}
\hfil}
\vskip 40pt
\line{\hfil {\bigrm P. L. White}\hfil}
\vskip 10pt
\line{\hfil {\it Physics Department, University of Southampton,}\hfil}
\line{\hfil {\it Southampton SO9 5NH, UK.}\hfil}
\vskip 40pt

\heading{\hfil Abstract }
In order that discrete symmetries should not be violated by gravitational
effects, it is necessary to gauge them. In this paper we discuss the
gauging of $\Z_N$ from the breaking of a high energy $SU(N)$ gauge symmetry,
and derive consistency conditions for the resulting discrete symmetry from
the requirement of anomaly cancellation in the parent symmetry. These
results are then applied to a detailed analysis of the possible discrete
symmetries forbidding proton decay in the minimal supersymmetric
standard model.

\vfill

\eject
\num=0
\sec=1
\heading{1 Introduction}
It has now been known for some time that discrete symmetries are likely to
be strongly violated by gravitational effects [\ref{\grav}]. This is
a problem for the minimal supersymmetric standard model (``MSSM''), where
the possible renormalisable interaction terms in the
Lagrangian are constrained not merely by Lorentz and gauge invariance, but
also by the requirement that a discrete symmetry, R parity, is not broken
by the Lagrangian. This symmetry is introduced purely in order to avoid
the presence of Yukawa
couplings which would lead to phenomenologically unacceptable proton decay,
and a number of other possible discrete symmetries have also been
studied [\ref{\Rparviol}].
In order to ensure that such symmetries are respected by
the gravitational interactions it is necessary to gauge them [\ref{\gds}],
[\ref{\AMW}]. Gauging a discrete symmetry means obtaining it
as the residual symmetry after the spontaneous breaking of some high energy
gauge symmetry.

Since this high energy symmetry must originally have been anomaly free, it
is possible to derive consistency conditions for the discrete symmetry, and
this has been done for the case of a $U(1)$ gauge symmetry breaking
to a $\Z_N$ discrete symmetry [\ref{\dscc}]. These results have been applied
to the possible symmetries of the MSSM [\ref{\IR}], with the result that
only a very restricted choice of such symmetries is possible.

In this paper we shall discuss another possibility for the production of
discrete symmetries, namely that of the breaking of $SU(N)$ to its $\Z_N$
centre, derive the resulting consistency conditions,
and apply our results to the MSSM. The structure of the paper is as follows.
Following this introduction, section 2 briefly discusses how this breaking
may occur in a non-trivial way. In section 3, we derive the consistency
conditions which must be obeyed by the $\Z_N$ charges of the theory. These
are then applied to the MSSM in section 4. Section 5 is the conclusion.

\num=0 \sec=2
\heading{2 Symmetry Breaking of $SU(N)$ to its Centre}
In this section we shall begin by briefly discussing how an $SU(N)$
symmetry can be broken without breaking its $\Z_N$ centre (consisting
of those matrices in $SU(N)$ of form
${exp}({2n\pi i\over N})\times {\bf 1} $
where {\bf 1} is the $N\times N$ unit matrix), an idea which was
mentioned in reference [\ref{\AMW}]. The simplest such mechanism is
that this breaking could occur
through a field in the adjoint representation acquiring a vacuum expectation
value (hereafter ``vev''), but it is also possible to break the $SU(N)$
by giving a vev to a field in any representation which has zero charge under
the $\Z_N$. The $\Z_N$ charge of a field which is in an irreducible
representation of $SU(N)$ with $n_u$ upper and $n_l$ lower vector indices is
given by $n_l-n_u$, and is defined modulo $N$. Although this gives many
possible breaking mechanisms, we shall only describe breaking through the
adjoint and $N$ index symmetric representations.

To break $SU(N)$ completely down to $\Z_N$ is trivial, and can be done
either by repeated
breaking or by choosing a representation of high enough dimension that when
it
gains a vev the
group breaks completely. However, for reasons which will become apparent
when we come to discuss consistency conditions which must be satisfied by
the discrete symmetry, this leads to extremely
tight constraints on the resulting $\Z_N$ and so is not the most
interesting case. Another possibility is that breaking occurs to give
part or all of the standard model gauge group
$SU(3)_C\times SU(2)_L\times U(1)_Y$, either directly from $SU(N)$ or with
some non-trivial mixing with other symmetry groups. This can occur in
many ways, given enough effort in constructing the potentials. As an
example we construct a model in which $N=3$ and $SU(3)\times U(1)$ breaks
to $SU(2)\times U(1) \times \Z_3$. From this it is obvious how one can
go about constructing more elaborate theories.

We begin with a gauge group $SU(3)\times U(1)_X$, which we break by giving
a vev to a scalar $A$ in the adjoint representation of zero $X$ charge. The
vev will be taken to be of form
$$<A>=\sqrt{3}aT_8=a \left( \matrix{ 1  & 0 & 0 \cr
                                     0  & 1 & 0 \cr
                                     0  & 0 & -2 \cr} \right)
\eqnlbl{\Avev} $$
where $a$ is a constant of mass dimension 1. Thus we have breaking of form
$$SU(3)\times U(1)_X\to SU(2)\times U(1)_8\times \Z_3 \times U(1)_X \eqnum$$
where $U(1)_8$ is the $U(1)$ symmetry generated by the $SU(3)$ generator
$T_8$. We now introduce a field $B_{ijk}$ in the three index symmetric
representation with non-zero $X$ charge, and give it a vev in the 333
component. This then leaves the full symmetry breaking
$$\eqalign{
SU(3)\times U(1)_X &\to SU(2)\times U(1)_8\times \Z_3 \times U(1)_X \cr
    &\to SU(2)\times U(1)_Y\times \Z_3 \cr
}\eqnum$$
Here $U(1)_Y$ is a linear combination of $U(1)_8$ and $U(1)_X$.
The potential which will give this breaking is
$$\eqalign{
V(A,B)=&\lambda_1 \bigl(\hbox{tr}(A^2)-6a^2\bigr)^2
        +\lambda_2\hbox{tr}\bigl( \vert (A+2a^{\prime})_i{}^mB_{mjk}
                                  +\hbox{cyclic}\vert^2\bigr) \cr
      &\quad  +\lambda_3\hbox{tr}\bigl(\vert B\vert^2-b^2\bigr)^2 \cr
}\eqnlbl{\potl} $$
Here $\lambda_i>0\ \ \forall\, i$, and at tree level we impose
$a=a^{\prime}$. It is simple to check that this
breaking is stable under (small) radiative corrections, even if the equality
$a=a^{\prime}$ is broken. Similarly,
choosing another form of the potential will allow us to break
$SU(2)\times U(1)_X$ to $U(1)$, as in the standard model.

\num=0 \sec=3
\heading{3 $\Z_N$ Consistency Conditions}
In this section we derive consistency conditions for the residual $\Z_N$
symmetry. These come from two sources: firstly the requirement that in the
original $SU(N)$ gauge theory there were no anomalies; and secondly the
observation that for each irreducible representation the total $\Z_N$
charge must be zero mod $N$. The latter condition will be much more
restrictive in the case where the breaking of $SU(N)$ is to $\Z_N$ only.
Since this discussion is rather elaborate, the reader who is uninterested
in the derivation is recommended to skip to the summary of results at the
end of the section.

We shall use Dynkin indices [\ref{\Dynk}] to describe
representations of $SU(N)$ (for reviews see [\ref{\gpthrev}]). For $SU(N)$
the Dynkin indices take the form of a set of $N-1$ non-negative integers
which we shall call $a_n$. Each distinct set of Dynkin indices corresponds
to an irreducible representation of $SU(N)$, and such properties as the
dimension and Casimir invariants of the representation can be represented
as functions of these indices. The Dynkin indices of the most common
representations of $SU(N)$ are then given in Table 1 below. It is clear
that for all the representations listed here the $\Z_N$ charge of a
particle in an irreducible representation with Dynkin indices
$a_n$ is given by the relation
$$Q=\sum_{p=1}^{N-1} pa_p \eqnlbl{\defQ} $$
and in fact it is easy to use the rules
for finding direct product of representations to check that this is true
for all irreducible representations of $SU(N)$.

\midinsert
$$\vbox {
\line{\qquad\qquad\bf Table 1\hfil}
\bigskip
\offinterlineskip \halign
{         \hbox{\vrule height 12pt depth 3pt width 0pt}#&
          \vrule#&  \hfil\quad\hbox{#}\quad\hfil&
          \vrule#&  \hfil\enskip\hbox{#}\enskip\hfil&
          \vrule#
          \cr
    \noalign{\hrule}

&& Representation && Dynkin indices $(a_1\ldots a_{N-1})$ & \cr
\noalign{\hrule}
&& singlet && $(0\ldots 0)$ & \cr
\noalign{\hrule}
&& fundamental && $(10\ldots 0)$ & \cr
\noalign{\hrule}
&& 2 index symmetric && $(20\ldots 0)$ & \cr
\noalign{\hrule}
&& 2 index anti-symmetric && $(010\ldots 0)$ & \cr
\noalign{\hrule}
&& $N-1$ index anti-symmetric && $(0\ldots 01)$ & \cr
\noalign{\hrule}
&& adjoint && $(10\ldots 01)$ & \cr
\noalign{\hrule}
\cr}} $$
\endinsert

We now wish to use the anomaly cancellation constraints in the form
of Dynkin indices. We firstly have the constraint that
$SU(N)\times SU(N)\times U(1)$ anomalies must cancel. This is equivalent to
the statement that
$$\sum_{\hbox{all reps $\Lambda$}} q_{\Lambda} I_2 (\Lambda) =0
\eqnum$$
where $q_{\Lambda}$ is the $U(1)$ charge of the fermions in the
representation $\Lambda$, and $I_2(\Lambda)$ is the second order index of
$\Lambda$, whose
relation to the second order Casimir invariant $C_2(\Lambda)$ is given by
$$I_2(\Lambda)=C_2(\Lambda)D(\Lambda) \eqnum$$
where $D(\Lambda)$ is the dimension of the representation. From this we
may use the standard results (using the notation of Slansky
[\ref{\gpthrev}])
that, in root space with $\delta$ equal to half the sum
of the positive roots,
$$C_2(\Lambda)=(\Lambda+2\delta,\Lambda) \eqnum$$
Here the symbol $\Lambda$ is used to represent the vector in weight space
corresponding to the highest weight of the representation $\Lambda$. This
may be expanded in terms of the Dynkin indices $a_n$ (which are the
components of $\Lambda$ in the Dynkin basis) to give for $SU(N)$
$$C_2(\Lambda)=\
\sum_{m=1}^{N-1} \bigl \{
N(N-m)ma_m +m(N-m)a_m^2 +\sum_{n=0}^{m-1}2n(N-m)a_na_m \bigr \}
\eqnlbl{\defCtwo} $$
and similarly
$$\eqalign{
D(\Lambda)=\, &\prod_{{\rm positive\ roots\ }\alpha}
              {(\Lambda+\delta,\alpha)\over(\delta,\alpha)} \cr
         =\,&\prod_{p=1}^{N-1}\Bigl [
     {1\over p!}\prod_{q=p}^{N-1}\bigl (\sum_{r=q-p+1}^p (1+a_r) \bigr )
                    \Bigr ]   \cr
      =\,& \qquad(1+a_1)(1+a_2)(1+a_3)\ldots (1+a_{N-1})\times\cr
&       \left ({2+a_1+a_2\over 2} \right )
        \left ({2+a_2+a_3\over 2} \right )
        \left ({2+a_3+a_4\over 2} \right )\ldots
        \left ({2+a_{N-2}+a_{N-1}\over 2} \right )\times\cr
&       \quad\left ({3+a_1+a_2+a_3\over 3} \right )\ldots
        \left ({3+a_{N-3}+a_{N-2}+a_{N-1}\over 3} \right )\times\cr
&       \qquad\ldots\times
        \left ({N-1+a_1+\ldots+a_{N-1})\over N-1} \right ) \cr
}\eqnlbl{\defD} $$

The corresponding equation for the cancellation of the
$SU(N)\times SU(N) \times SU(N)$ anomaly is that the object $I_3(\Lambda)$
should vanish, where $I_3(\Lambda)$ is given by $C_3(\Lambda)D(\Lambda)$.
The expression for $C_3(\Lambda)$ in terms of the Dynkin indices is
[\ref{\BG}]
$$
C_3(\Lambda)=\sum_{p,q,r=1}^{N-1} d_{pqr}(a_p+1)(a_q+1)(a_r+1)
\eqnlbl{\defCthree} $$
with the totally symmetric tensor $d_{pqr}$ defined by
$$d_{pqr}:=\half p(N-2q)(N-r) \qquad \hbox{where }p\le q\le r
\eqnlbl{\defd} $$
Note that we have changed both the notation and the normalisation of
reference [\ref{\BG}], since the normalisation is irrelevant for
our purposes, and the factor half in \defd\ will simplify later algebra.

We now wish to find how these conditions constrain the possible $\Z_N$
symmetries of the theory. In order to do this, we shall need to derive
a number of relations involving the Dynkin indices. We first
note that $D(\Lambda)=D(a_1\ldots a_{N-1})$ is obviously
an integer (since it is the dimension of a representation).
Similarly, $D(a_1\ldots a_{N})$ is defined as the dimension of a
representation of $SU(N+1)$
and so is also an integer, where $a_m$ is taken to be the same in both
cases for $m<N$, and $a_N$ is arbitrary. We may
now use equation \defD\ to find that
$$\eqalign{
D(a_1\ldots a_N)=& (1+a_N)\left({2+a_{N-1}+a_N\over 2}\right )\times \ldots
       \times \left ( {N+a_1+\ldots+a_N \over N} \right )
        D(a_1\ldots a_{N-1})        \cr
     =&{1\over N!}\left [\prod_{m=1}^N(m+a_{N-m+1}+\ldots+a_N)\right ]
                         D(a_1\ldots a_{N-1})\cr
}\eqnum$$
A useful definition is then $f_0(a_1\ldots a_N)$ from
$$D(a_1\ldots a_N)=:{1\over N}f_0(a_1\ldots a_N)D(a_1\ldots a_{N-1})
\eqnlbl{\deff} $$
Now, since $D(a_1\ldots a_N)$ is an integer, it is clear that
$$f_0(a_1\ldots a_N)D(a_1\ldots a_{N-1})=0 \qquad \hbox{mod $N$} \eqnum$$
This equation is true for all $a_N$ (so long as $a_N$ is a non-negative
integer) and so it is also true if we replace
$a_N$ by $(a_N+1)$. Thus we define
$$\eqalign{
f_{i+1}(a_N)=&f_i(a_N+1)-f_i(a_N) \cr
            =&{\partial f_i(a_N)\over \partial a_N}
             +{1\over 2!}{\partial^2 f_i(a_N)\over \partial a_N^2}
             +{1\over 3!}{\partial^3 f_i(a_N)\over \partial a_N^3}
             +\ldots     \cr
}\eqnlbl{\fidef} $$
from which it is clear that
$$f_i(a_1\ldots a_N)D(a_1\ldots a_{N-1})=0
        \qquad \hbox{mod $N$}\quad\forall\ i
\eqnlbl{\fcond} $$
In particular we have, suppressing all the $a_m$ dependence except that
on $a_N$,
$$\eqalignno{
f_N(a_N)=&{\partial^N f_0(a_N)\over \partial a_N^N} &\eqalnum\cr
f_{N-1}(a_N)=&{\partial^{N-1} f_0(a_N)\over \partial a_N^{N-1}}
           +{(N-1)\over 2}{\partial^N f_0(a_N)\over \partial a_N^N}
             &\eqalnum\cr
f_{N-2}(a_N)=&{\partial^{N-2} f_0(a_N)\over \partial a_N^{N-2}}
       +{(N-2)\over 2}{\partial^{N-1} f_0(a_N)\over \partial a_N^{N-1}} \cr
      &\qquad +\left({(N-2)\over 3!}+{(N-2)(N-3)\over 8}\right)
                 {\partial^N f_0(a_N)\over \partial a_N^N}
                 & \eqalnum
}$$
These equations may be derived using \fidef\ and remembering that
$$
{\partial^{N+1} f_0(a_N)\over \partial a_N^{N+1}}=0
\eqnum $$

We may now expand out the explicit form of these three equations, and we
discover that after some simplification
$$\eqalignno{
f_N(a_1\ldots a_N)=& N!& \eqalnumlbl{\deffn} \cr
f_{N-1}(a_1\ldots a_{N-1}0)=& \sum_p pa_p& \eqalnumlbl{\deffone} \cr
f_{N-2}(a_1\ldots a_{N-1}0)=&{1\over 2(N-1)}\sum_p \left(
        pa_p^2-p^2a_p+2\sum_{q<p}qa_qa_p \right ) &\eqalnumlbl{\defftwo} \cr
}$$

We must now use \defCtwo\ and \defCthree, together with \deffone\ and
\defftwo\ to derive constraints on the $\Z_N$ charges of the theory. We
begin with the requirement that the $SU(N)\times SU(N)\times U(1)$ anomaly
cancels. The contribution to this anomaly of a representation $\Lambda$
is $q_{\Lambda}C_2(\Lambda)D(\Lambda)$
which can be expanded out with \defCtwo\ to give
$$\eqalign{
q_{\Lambda}C_2(\Lambda)D(\Lambda)=&-q_{\Lambda}(\sum_p pa_p)^2D(\Lambda)
      +q_{\Lambda}N^2D(\Lambda)(\sum_p a_p) \cr
     &+q_{\Lambda}ND(\Lambda)\Bigl (\sum_{p=1}^{N-1}
                    (-p^2a_p +pa_p^2 +\sum_{q<p}2qa_qa_p) \Bigr)  \cr
}\eqnum$$
If we now assume that the $U(1)$ charges have been normalised so that
$q_{\Lambda}$ is an integer, we have a contribution to the anomaly
$$\sum_{\Lambda}q_{\Lambda}C_2(\Lambda)D(\Lambda)
       =-\sum_{\Lambda}q_{\Lambda}(\sum_p pa_p)^2D(\Lambda)
       \qquad\hbox{mod $N^2$}
\eqnum$$
where we have used \defftwo\ and \fcond\ in showing that the extra terms
are zero mod $N^2$. Now for each irreducible representation $\Lambda$ there
are $D(\Lambda)$ particles of charge $(\sum_p pa_p)$, and so we see that
the requirement that the anomaly vanish can be expressed as
$$
\sum_i q_i Q_i^2 = \, 0 \qquad \hbox {mod $N^2$}
\eqnlbl{\ctwo} $$
where the sum is over all particles labelled $i$ of $U(1)$ and $\Z_N$ charge
$q_i$ and $Q_i$ respectively.

In the case where the $U(1)$ symmetry of the standard model arises as an
unbroken subgroup of the $SU(N)$, or as a linear combination of such a
subgroup and another $U(1)$ symmetry, say $U(1)^{\prime}$, we can use the
traceless of $SU(N)$ to show that for particles $i$ in each irreducible
representation with charge $q_i$ of which $q_i^{\prime}$ comes from
$U(1)^{\prime}$
$$\sum_i q_i =\sum_i q_i^{\prime} \eqnum$$
and so \ctwo\ is unchanged.

The consistency condition for the cancellation of the $SU(N)^3$ anomaly is
derived in much the same way, and after a certain amount of algebra we
find that
$$\eqalign{
C_3(\Lambda)=& (\sum_pa_p)^3
 +\sum_p \quarter N^2 a_p p(N^2-3Np+2p^2)
 +\sum_p{3N\over 2}a_p^2
    (p^3-{\scriptstyle{3\over 2}}Np^2+\half N^2p)\cr
&+\sum_{q>p}{3N\over 2}a_pa_q(p^2q+q^2p-Np^2-2Npq+N^2p)  \cr
&+\sum_p a_p^3(-{\scriptstyle{3\over 2}}Np^2+\half N^2p)
 +\sum_{q>p} a_pa_q^2(-{\scriptstyle{9\over 2}}Npq
                 +{\scriptstyle{3\over 2}}N^2p)\cr
& +\sum_{q>p} a_p^2a_q(-3Np^2-{\scriptstyle{3\over 2}}Npq
                 +{\scriptstyle{3\over 2}}N^2p)
 +\sum_{p<q<r}a_pa_qa_r(-6Npq-3Npr+3N^2p) \cr
}\eqnum$$
Extensive use of \deffone, \defftwo, and \fcond\ reduces this to
$$C_3(\Lambda)D(\Lambda)=(\sum_p pa_p)^3D(\Lambda)
   \qquad\hbox{mod $N^2$}
\eqnum$$
from which we conclude that the anomaly cancellation condition is that
$$\sum_i Q_i^3 = \, 0 \qquad \hbox {mod $N^2$}
\eqnlbl{\cthree} $$

It should be noted that we are assuming that the $\Z_N$ charge $Q$ of a
particle in the representation labelled by $a_p$ is given by $\sum_ppa_p$,
although this is in fact only true
mod $N$ and some of our equations require definitions to be valid
mod $N^2$. However, it is easy to check that it does not matter if we
select another value for $Q$ (equal to the first mod $N$) so long as we
pick the same
value for all fermions in each representation. In practical terms
this amounts to selecting one set of numbers to represent the $\Z_N$
charges and sticking to it. We shall use the simplest such set,
namely $\{ -{N-1\over 2},\ldots, {N-1\over 2} \}$ for $N$ odd, and
$\{ -{N\over 2}+1,\ldots, {N\over 2} \}$ for $N$ even.

A further problem is that the $U(1)$ charges are assumed integer. If they
are not, then it is necessary to normalise them by multiplying by some
overall constant until they are.

In addition to \ctwo\ and \cthree, we can use \deffone\ to obtain the result
that for each irreducible representation separately
$$\sum_i Q_i =\, 0 \qquad \hbox{mod $N$} \eqnlbl{\cone} $$
Thus if the $SU(N)$ symmetry breaks trivially to $\Z_N$ only, and
thus has no mixing with the gauge group of the standard model (or their
ancestors) at high energies, this constraint can be applied to each set
of fermions with the same quantum numbers (including $\Z_N$ charge)
separately (since before
$SU(N)$ breaking they must have belonged to different irreducible
representations) and so we shall discover that our theory is constrained
effectively only to generation symmetries. This  however does not apply
in the case where we allow a more complicated breaking, and then we may
only impose that \cone\ is satisfied if the sum is over all fermions
which might originally have been in the same representation, that is
those with the same $\Z_N$ charge.

These constraints ignore the possible effects of particles which are not
visible at low energies because they have acquired large masses
[\ref{\dscc}].
This can occur through a fermion gaining a Majorana mass, which is only
possible if it has $\Z_N$ charge 0 or ${N\over 2}$ (the latter only for
even $N$ when $Q_i=Q_j={N\over 2}$;
we are here using our convention for the $\Z_N$ charges which gives
that $-{N \over 2}<Q_i\le {N\over 2}$ ) and no $U(1)$ charge. Fermions of
$\Z_N$ charge 0 do not matter
for the consistency conditions, and so the only effect is for $N$ even
when each such fermion adds ${N^3\over 8}$ to the right hand side of
equation \cthree, and removes the constraint \cone\ for charge
${N\over 2}$.

It is also possible for two fermions, say $i$ and $j$, to combine to
obtain a Dirac mass. This can only occur if $Q_i+Q_j=0$ or
$Q_i+Q_j=N$ , and if $q_i+q_j=0$ (remember that we must not violate
the $U(1)$ symmetry).
Thus we have two effects. Firstly, the constraint \cone\ is weakened so
that the sum now runs not over all fermions of the same $\Z_N$
charge $Q$ but over all fermions of charge $Q$ and $-Q$. The case
where the two fermions both have charge ${N\over 2}$ does not give
any further restriction.

We have now finished the derivation of the consistency conditions, and
will thus summarise them below. For $N$ odd we have
$$\eqalignno{
Q(n_Q-n_{-Q}) =&\, 0 \qquad \hbox{mod $N$ $\forall$ Q} &
        \eqalnumlbl{\coneNodd} \cr
\sum_i q_iQ_i^2 =&\, 0 \qquad \hbox{mod $N^2$} & \eqalnumlbl{\ctwoNodd} \cr
\sum_i Q_i^3 =&\, 0 \qquad \hbox{mod $N^2$} & \eqalnumlbl{\cthreeNodd} \cr
}$$
while for $N$ even these become
$$\eqalignno{
Q(n_Q-n_{-Q}) =&\, 0 \qquad \hbox{mod $N$ }\forall\ Q\ne {N\over 2} &
       \eqalnumlbl{\coneNeven} \cr
\sum_i q_iQ_i^2 =& \, 0 \qquad \hbox{mod $N^2$} &
       \eqalnumlbl{\ctwoNeven} \cr
\sum_i Q_i^3 =& \, \eta\left({N\over2}\right)^3
   \qquad\hbox{mod $N^2$}
              & \eqalnumlbl{\cthreeNeven} \cr
}$$
In all of these equations the sum over $i$ is the sum over all fermions
labelled $i$ whose $\Z_N$ charge is $Q_i$, $n_Q$ is the number of fermions
of charge
$Q$, and $\eta$ is an arbitrary integer. Note that for any given particle
content, these are necessary but not sufficient conditions that the $\Z_N$
symmetry might have originated in an anomaly free high energy $SU(N)$
symmetry.

\num=0 \sec=4
\heading{4 Application to the MSSM}
In this section we demonstrate the use of our results by applying them to
the specific case of the MSSM. Here we are severely restricted in which
$\Z_N$ symmetries are acceptable from the requirement that terms in the
Lagrangian leading to excessive proton decay be forbidden. The first part
of this section will be essentially a summary of the analysis in
[\ref{\IR}])

We shall impose a number of restrictions on the possible symmetries which
we consider.
These are that the symmetries be generation-blind with three generations;
that the only singlet
be the right-handed neutrino; and that $N\le 4$. The second of these is
particularly restrictive, since the inclusion of particles which interact
only at high energy through the $SU(N)$ symmetry, and so only affect the
theory below the breaking scale through the $\Z_N$, can seriously
weaken the constraints. This effect is more pronounced than the
case where breaking from $U(1)$ is considered, since then more of the
constraints involve mixing with other symmetries, which singlets of course
cannot affect. Including such particles is rather
complicated, since the exact mechanism by which $SU(N)$ breaks will
affect how many such particles are allowed. For example, if $SU(N)$
breaks to give the $SU(2)$ of the standard model, then some of the
singlets will have originally come from $SU(N)$ multiplets which after
breaking give the $SU(2)$ multiplets of the standard model.
It is not clear how
one might handle such a situation in general, although it would be possible
to analyse it for each particular case.

The MSSM, with the inclusion of right-handed neutrinos but no other singlets,
has the gauge group $SU(3)\times SU(2)\times U(1)$, and particle content
(with gauge couplings indicated):

$$\vbox {
\offinterlineskip \halign
{         \hbox{\vrule height 12pt depth 3pt width 0pt}#&
          $\ #\ $ & $#\ $ &
          $\qquad\qquad\ #\ $ & $#\ $
         \cr
& q & (3,2,{\scriptstyle{1\over 6}}) & L & (1,2,-\half)  \cr
& d & (\bar 3,1,\third) & e & (1,1,1)  \cr
& u & (\bar 3,1,-{\scriptstyle{2\over 3}}) & \nu & (1,1,0) \cr
& H & (1,2,-\half) & \bar H & (1,2,\half)  \cr
\cr}}\eqnum$$

The gauge symmetry allows a number of Yukawa couplings, including
the dimension four terms $LHe$, $qHd$, $q\bar Hu$, and $\mu H\bar H$
all of which must be permitted by the discrete symmetry in order to
obtain the correct structure of the standard model, although with the
inclusion of extra singlets it is possible to avoid the necessity
of allowing $\mu H\bar H$. In addition to
these, there is the neutrino mass term $L\bar H\nu$ which is an acceptable
addition to the standard model (although it is interesting that it
is possible to construct models where the discrete symmetry prevents
neutrino masses), and terms allowing proton decay. These
are lepton number violating terms
$$LLe \qquad\qquad LQd  \eqnlbl{\Lviol} $$
and the baryon number violating term
$$udd. \eqnlbl{\Bviol} $$
We ignore
a possible lepton number violating term $\mu L\bar H$ because with our
assumption that the term $\mu H\bar H$ is allowed it can always be removed
by a field redefinition if it appears.
Although one conventionally requires all of these to be banned by the
discrete symmetry, it is in fact sufficient that either \Lviol\ or \Bviol\
should be banned together with certain higher dimension terms. We shall
adopt the view that to be acceptable, a symmetry must either prevent all
of \Lviol\ and \Bviol\ or else must ban all lepton or baryon number violating
terms of dimension four or five.

In terms of the $\Z_N$ charges, the requirement that none of the
standard model Yukawa interactions violate the symmetry is
$$\eqalign{
Q(H)+Q(\bar H)=&\, 0 \quad\hbox{mod $N$} \cr
Q(L)+Q(H)+Q(e)=&\, 0 \quad\hbox{mod $N$} \cr
Q(q)+Q(H)+Q(d)=&\, 0 \quad\hbox{mod $N$} \cr
Q(q)+Q(\bar H)+Q(u)=&\, 0 \quad\hbox{mod $N$} \cr
}\eqnlbl{\SMYuk} $$

We now begin by considering the specific case $N=2$, where our $\Z_2$
symmetry is a relic of a high energy $SU(2)$ symmetry. It is clear that
for this case the constraints \coneNeven\ and \cthreeNeven\ are trivial
(of course we would expect the latter, since there are no $SU(2)^3$
anomalies). \ctwoNeven\ gives the constraint
$$6Q(q)^2+6Q(d)^2-18Q(L)^2+18Q(e)^2-3Q(H)^2+3Q(\bar H)^2=0
        \qquad\hbox{mod 4} \eqnum$$
since we must multiply all the charges by 6 to make them integers, and
there are three generations. After some rearrangement, and using
\SMYuk\ and the fact that each charge can only take the values 0 or 1,
we find that this constraint is always satisfied. Thus we find that
any $\Z_2$ symmetry which bans all the unwanted terms in the Lagrangian is
acceptable from the point of view of anomaly contraints. Examples of
typical theories from which such a symmetry could appear are the wide range
of theories involving $SU(2)_R$ (where the subscript $R$ means
``right''), and in fact the usual R-parity could be such a case. To see
this, note that all discrete symmetries
added to the standard model can only really be defined modulo the discrete
symmetries which are already there and gauged, and these are the $\Z_2$
centre of $SU(2)_L$, the $\Z_3$ centre of $SU(3)_C$, and discrete subgroups
of $U(1)_Y$. By this argument, the only inequivalent $\Z_2$
symmetries are $\Z_{2L}$ (from $SU(2)_L$), R-parity, and lepton number,
although it is not easy to see whether it is possible to construct an
$SU(2)$ symmetry which breaks to give the last.

The case where $N=3$ is less trivial. The simplest way of solving for all
the possible discrete symmetries is to write down every possible set of
charges satisfying \SMYuk\ and then to test them all against the consistency
conditions. This can be done by noticing that given \SMYuk\ we need only
assign $\Z_3$ charges to $q$, $L$, $\nu$, and $H$ to define the symmetry
completely, and $\nu$ will usually be given by the consistency
conditions. We thus
only have $3^3$ symmetries to consider, and many of these are equivalent
to one another under reflections (replacing each charge $Q$ with $-Q$).
The constraints \coneNodd\ to \cthreeNodd\ are then entirely described
by
$$\eqalign{
Q(H)+Q(\bar H)\ =&\ 0\cr
Q(q)^2-2Q(u)^2+Q(d)^2\ =&\ 0  \cr
2Q(q)+Q(d)+Q(u)+2Q(L)+Q(e)+Q(\nu)\ =&\ 0 \qquad\hbox{mod 3} \cr
}\eqnum$$
where we have simplified using the fact that our charges are all in the
set $\{ -1,0,+1 \}$.
This gives a fairly straightforward result, and rather than
going through all the algebra in detail we shall simply give the
possible symmetries, which are listed in Table 2. Note that because
we have 3 generations,
the consistency conditions are much weaker than we might have expected.

\midinsert
$$\vbox {
\line{\qquad\qquad\bf Table 2: $\Z_3$ charges consistent with the MSSM\hfil}
\bigskip
\offinterlineskip \halign
{         \hbox{\vrule height 12pt depth 3pt width 0pt}#&
          \vrule#&  \hfil\enskip\hbox{#}\enskip\hfil&
          \vrule#&  \hfil\enskip\hbox{$#$}\enskip\hfil&
          \vrule#&  \hfil\enskip\hbox{$#$}\enskip\hfil&
          \vrule#&  \hfil\enskip\hbox{$#$}\enskip\hfil&
          \vrule#&  \hfil\enskip\hbox{$#$}\enskip\hfil&
          \vrule#&  \hfil\enskip\hbox{$#$}\enskip\hfil&
          \vrule#&  \hfil\enskip\hbox{$#$}\enskip\hfil&
          \vrule#&  \hfil\enskip\hbox{$#$}\enskip\hfil&
          \vrule#&  \hfil\enskip\hbox{$#$}\enskip\hfil&
          \vrule#
          \cr
    \noalign{\hrule}

&& Symmetry && Q(H) && Q(\bar H) && Q(q) && Q(u) && Q(d)
   && Q(L) && Q(e) && Q(\nu) & \cr
\noalign{\hrule}
&& A && 0 && 0 && 0 && 0 && 0 && 1 && -1 && -1 & \cr
\noalign{\hrule}
&& B && 0 && 0 && 1 && -1 && -1 && 0 && 0 && 0 & \cr
\noalign{\hrule}
&& C && 0 && 0 && 1 && -1 && -1 && 1 && -1 && -1 & \cr
\noalign{\hrule}
&& D && 0 && 0 && 1 && -1 && -1 && -1 && 1 && 1 & \cr
\noalign{\hrule}
\cr}} $$
\endinsert

Having found all the possible symmetries, we now find whether they give
acceptable constraints on the MSSM Lagrangian. B clearly does not
prevent any of the proton decay terms and is of no interest to us.
It is in fact the $\Z_3$ centre of the colour group.
None of these symmetries bans both \Bviol\ and \Lviol, but A,C, and
D will protect lepton number, and they are all equivalent modulo
$\Z_3$ colour to the residual $\Z_3$ from a generational symmetry.

Thus, there is only one possible anomaly free symmetry for $N=3$
which will protect the proton from excessive decay, although it is
noticable that this analysis would not give any possibilities if we had
not included right handed neutrinos, and that
there is no reason here why such neutrinos should not gain a Dirac mass.

We conclude our analysis by studying the case where $N=4$. This again gives
fairly simple results, listed in Table 3.
\midinsert
$$\vbox {
\line{\qquad\qquad\bf Table 3: $\Z_4$ charges consistent with the MSSM\hfil}
\bigskip
\offinterlineskip \halign
{         \hbox{\vrule height 12pt depth 3pt width 0pt}#&
          \vrule#&  \hfil\enskip\hbox{#}\enskip\hfil&
          \vrule#&  \hfil\enskip\hbox{$#$}\enskip\hfil&
          \vrule#&  \hfil\enskip\hbox{$#$}\enskip\hfil&
          \vrule#&  \hfil\enskip\hbox{$#$}\enskip\hfil&
          \vrule#&  \hfil\enskip\hbox{$#$}\enskip\hfil&
          \vrule#&  \hfil\enskip\hbox{$#$}\enskip\hfil&
          \vrule#&  \hfil\enskip\hbox{$#$}\enskip\hfil&
          \vrule#&  \hfil\enskip\hbox{$#$}\enskip\hfil&
          \vrule#&  \hfil\enskip\hbox{$#$}\enskip\hfil&
          \vrule#
          \cr
    \noalign{\hrule}

&& Symmetry && Q(H) && Q(\bar H) && Q(q) && Q(u) && Q(d)
   && Q(L) && Q(e) && Q(\nu) & \cr
\noalign{\hrule}
&& A && 0 && 0 && 0 && 0 && 0 && 1 && -1 && -1 & \cr
\noalign{\hrule}
&& B && 0 && 0 && 1 && -1 && -1 && 0 && 0 && 0\ or\ 2 & \cr
\noalign{\hrule}
&& C && 0 && 0 && 1 && -1 && -1 && 1 && -1 && -1 & \cr
\noalign{\hrule}
&& D && 0 && 0 && 1 && -1 && -1 && 2 && 2 && 0\ or\ 2 & \cr
\noalign{\hrule}
&& E && 0 && 0 && 1 && -1 && -1 && -1 && 1 && 1 & \cr
\noalign{\hrule}
&& F && 0 && 0 && 2 && 2 && 2 && 1 && -1 && -1 & \cr
\noalign{\hrule}
&& G && 1 && -1 && 1 && 0 && 2 && 0 && -1 && -1 & \cr
\noalign{\hrule}
&& H && 1 && -1 && -1 && 2 && 0 && 2 && 1 && 1 & \cr
\noalign{\hrule}
&& I && 2 && 2 && 1 && 1 && 1 && 1 && 1 && 1 & \cr
\noalign{\hrule}
&& J && 2 && 2 && 1 && 1 && 1 && -1 && -1 && -1 & \cr
\noalign{\hrule}
\cr}} $$
\endinsert

Note that we have ignored cases where all the charges are zero mod 2, since
such symmetries are merely $\Z_2$ symmetries. In general a $\Z_N$
symmetry which is a subgroup
of $\Z_M$ (so that $M>N$) obeys less restrictive conditions if it is
obtained
from a breaking of $SU(M)$ than from $SU(N)$. In this case if we
impose the standard model constraints \SMYuk\ and require that the $\Z_4$
charges are all either 0 or 2 (so as to give a $\Z_2$ symmetry), then the
consistency conditions are all trivially
satisfied, just as for $\Z_2$.

{}From the table there are seven possible $\Z_4$ symmetries which prevent all
the unwanted dimension four terms in the lagrangian, and three more which
preserve either lepton or baryon number but not both (although two
of those listed are in fact redundant, as they are products of others given
here). These symmetries also
include some which could prevent the neutrino gaining a large Dirac mass.

\num=0 \sec=5
\heading{5 Conclusion}
We have thus discussed how it is possible to break $SU(N)$ to smaller
gauge groups in such a way as to leave its $\Z_N$ centre as a residual
symmetry. The requirement that the high energy $SU(N)$ theory be anomaly
free then gives restrictive constraints on the resulting discrete symmetry,
although these are significantly weakened if we allow the addition to
the model of singlets which interact only through the $SU(N)$ and thus
carry non-zero charge under $\Z_N$.

These constraints can then be simply solved in the case of the MSSM for
$N\le4$ with the only singlet being the right-handed neutrino, to find
that while
$\Z_2$ symmetries are not restricted at all,
only a few possible
$\Z_3$ and $\Z_4$ symmetries are left.

While the constraints given here can be quite restrictive, it may well be
possible to derive further ones from a more detailed analysis of
the possible symmetry breaking mechanisms. For example, if $SU(3)_C$
arises from the breaking of $SU(N)$, then each irreducible representation
of $SU(N)$ must contain an $SU(3)_C$ multiplet, and so extra singlets can
only be introduced in conjunction with $SU(3)_C$ multiplets.

There are a number of interesting possibilities for getting discrete
symmetries from $SU(N)$ which we have not mentioned. Apart from the most
obvious cases of breaking the $SU(N)$ to $SU(M)$ (for $M<N$) or to $U(1)$,
and then breaking the latter to a discrete symmetry as usual,
we might consider the possibility that after breaking one might have two
distinct discrete symmetries. This is essentially covered by the usual
analysis, except that if we have both abelian
and non-abelian symmetries which break to discrete remnants $\Z_M$ and
$\Z_N$ respectively, then we will
have a mixed anomaly cancellation constraint of form
$$\sum_i q_iQ_i^2 =0  \qquad\hbox{gcd($N^2$,$MN$)} \eqnum$$
where the $\Z_M$ and $\Z_N$ charges are $q_i$ and $Q_i$, and gcd($N^2$,$MN$)
is the greatest common divisor of $N^2$ and $MN$. This assumes that we
select the $\Z_M$ charges to lie in a unique set, as for the $\Z_N$ ones.

Finally, we should  mention that there is not any particular reason why
discrete symmetries must not be non-abelian, although it is not easy to
see how the breaking might occur to give rise to these, and there is no
use for such symmetries in the standard model.

\heading{Acknowledgements}
I would like to thank Robert Foot, Ron King, Steven King, and Douglas Ross
for various helpful suggestions.

\heading{References}
\parindent -15pt
[\ref{\grav}] S.Hawking; {\it Phys. Lett.} {\bf 195B} (1987) 337

\quad G.V.Lavrelashvili, V.Rubakov, P.Tinyakov; {\it JETP Lett.}
              {\bf 46} (1987) 167

\quad S.Giddings, A.Strominger; {\it Nucl. Phys.} {\bf B306} (1988) 890

\quad S.Coleman; {\it Nucl. Phys.} {\bf B310} (1988) 643

[\ref{\Rparviol}] L.Hall, M.Suzuki; {\it Nucl. Phys.} {\bf B231} (1984) 419

\quad F.Zwirner; {\it Phys. Lett.} {\bf 132B} (1983) 103

\quad R.Barbieri, A.Masiero; {\it Nucl. Phys.} {\bf B267} (1986) 679

\quad I.Lee; {\it Nucl. Phys.} {\bf B246} (1984) 120

\quad J.Ellis, G.Gelmini, C.Jarlskog, G.Ross, J.Valle;
          {\it Phys. Lett.} {\bf 150B} (1985) 142

[\ref{\gds}] L.Krauss, F.Wilczek; {\it Phys. Rev. Lett.}
                {\bf 62} (1989) 1221

\quad T.Banks; {\it Nucl. Phys.} {\bf B323} (1989) 90

\quad L.Krauss; {\it Gen. Rel. Grav.} {\bf 22} (1990) 253

\quad J.Preskill, L.Krauss; {\it Nucl. Phys.} {\bf B341} (1990) 50

[\ref{\AMW}] M.Alford, J.March-Russell, F.Wilczek; {\it Nucl. Phys.}
            {\bf B337} (1990) 695

[\ref{\dscc}] L.E.Ibanez, G.G.Ross; {\it Phys. Lett.} {\bf 260B} (1991) 291

\quad T.Banks; {\it Phys. Rev.} {\bf D45} (1992) 424

\quad L.E.Ibanez, G.G.Ross; preprint CERN-TH.6000/91

[\ref{\IR}] L.E.Ibanez, G.G.Ross; {\it Nucl. Phys.} {\bf B368} (1992) 3

[\ref{\Dynk}] E.Dynkin; {\it Am. Math. Soc. Trans., Ser.2}
                     {\bf 6} (1957) 111, 245

[\ref{\gpthrev}] R.Slansky; {\it Phys. Rep} {\bf 79} (1981) 1

\quad R.Gilmore; ``{\it Lie Groups, Lie Algebras, and some of their
        Representations}''; (Wiley, New York, 1974)

\quad L.O'Raifeartaigh; ``{\it Group Structure of Gauge Theories}'';
        (Cambridge University Press, 1986)

[\ref{\BG}] J.Banks, H.Georgi; {\it Phys. Rev.} {\bf D14} (1976) 1159

\bye